\documentclass[pra,showpacs,twocolumn,aps,superscriptaddress,a4paper]{revtex4-1}
\usepackage{dcolumn,amssymb,amsmath,amsfonts,graphicx,latexsym,color}
\usepackage{epstopdf}
\usepackage{CJK}
\begin{document}

\begin{CJK*}{GBK}{song}

\title{Width of the confinement-induced resonance in a quasi-one-dimensional trap with transverse anisotropy}

\author{Fang Qin}
\email{qinfang@ustc.edu.cn}
\affiliation{Key Laboratory of Quantum Information, University of Science and Technology of China, Chinese Academy of Sciences, Hefei, Anhui 230026, P.R. China}
\affiliation{Synergetic Innovation Center of Quantum Information and Quantum Physics, University of Science and Technology of China, Hefei, Anhui 230026, P.R. China}
\author{Jian-Song Pan}
\email{panjsong@mail.ustc.edu.cn}
\affiliation{Key Laboratory of Quantum Information, University of Science and Technology of China, Chinese Academy of Sciences, Hefei, Anhui 230026, P.R. China}
\affiliation{Synergetic Innovation Center of Quantum Information and Quantum Physics, University of Science and Technology of China, Hefei, Anhui 230026, P.R. China}
\affiliation{Wilczek Quantum Center, School of Physics and Astronomy and T. D. Lee Institute, Shanghai Jiao Tong University, Shanghai 200240, P.R. China}
\author{Su Wang}
\email{james016@mail.ustc.edu.cn}
\affiliation{Key Laboratory of Quantum Information, University of Science and Technology of China, Chinese Academy of Sciences, Hefei, Anhui 230026, P.R. China}
\affiliation{Synergetic Innovation Center of Quantum Information and Quantum Physics, University of Science and Technology of China, Hefei, Anhui 230026, P.R. China}
\author{Guang-Can Guo}
\affiliation{Key Laboratory of Quantum Information, University of Science and Technology of China, Chinese Academy of Sciences, Hefei, Anhui 230026, P.R. China}
\affiliation{Synergetic Innovation Center of Quantum Information and Quantum Physics, University of Science and Technology of China, Hefei, Anhui 230026, P.R. China}

\date{\today}

\begin{abstract}
We theoretically study the width of the $s$-wave confinement-induced resonance (CIR) in quasi-one-dimensional atomic gases under tunable transversely anisotropic confinement. We find that the width of the CIR can be tuned by varying the transverse anisotropy. The change in the width of the CIR can manifest itself in the position of the discontinuity in the interaction energy density, which can be probed experimentally.
\end{abstract}


\maketitle

\section{Introduction}

Low-dimensional quantum gas has attracted more and more attentions since its experimental realization~\cite{CIR2010experiment,CIR2010review,Q2Dexp2010,Q2Dexp2011,Q2Dexp2014,Q2Dexp2015}.
Experimentally, quasi-low-dimensional quantum gases can be realized by introducing tight confinements with optical lattices to freeze one or two spatial degrees of freedom.
Besides, the interactions between two atoms can be tuned by a Feshbach resonance.
Based on these experimental techniques, the quasi-low-dimensional quantum gases can provide an ideal platform to study the fundamental many-body physics which is very different from that of three-dimensional gases.

The quasi-one-dimensional (quasi-1D) quantum gases can be realized by introducing a two-dimensional deep optical lattice in the transverse plane.
In this situation, the kinetic energy of the particles is too weak to drive them to the transversal excited energy levels.
When the two-body interaction cannot support a bound state, the transverse confinement can be used to modify scattering
along the unconfined direction by providing an additional structure of the transverse energy levels.
Varying the trapping frequencies, the effective one-dimensional scattering length even becomes divergent,
which is the so-called confinement-induced resonance (CIR)~\cite{Olshanii1998,Olshanii2003,Naidon2007,Peng2010,Zhang2011,Cui2012,kmatrix2004,kmatrixlwave2012,kmatrixlwave2014,kmatrixlwave2015,kmatrixlwave2016,kmatrixMelezhik2016,dipolarCIR2013,dipolarCIR20141,dipolarCIR20142,Lange2009,Melezhik2012,Melezhik2015,Melezhik2008,Melezhik2011,dwave2011,dwave2012,Peng2011,Qi2013}.

The transverse anisotropy is a basic tunable parameter for a quasi-1D quantum gas.
The CIR for an ultracold quasi-1D gas has been observed by measuring the atom loss and heating
rate with respect to different transversely anisotropic confinements~\cite{CIR2010experiment}.
The corresponding resonance position of the CIR has been theoretically studied using
the zero-energy $s$-wave-scattering pseudopotential approach~\cite{Peng2010,Zhang2011,Cui2012}.
With the $K$-matrix approach~\cite{kmatrix2004,kmatrixlwave2012,kmatrixlwave2014,kmatrixlwave2015}, the positions of $l$-wave CIRs in both isotropic and anisotropic harmonic waveguides have been calculated~\cite{kmatrixlwave2016,kmatrixMelezhik2016}.
Moreover, the positions of dipolar CIRs have been discussed in Refs.~\cite{dipolarCIR2013,dipolarCIR20141,dipolarCIR20142}.
As one basic quantity to describe a Feshbach resonance, the width of a CIR also attracts many attentions.
For example,
with the two-channel model of Lange et al.~\cite{Lange2009}, the shifts and widths of $s$- and $p$-wave CIRs in quasi-$1D$ isotropic traps have been studied~\cite{Melezhik2012,Melezhik2015}.
With the general grid method suggested in Ref.~\cite{Melezhik2008}, the impact of multichannel scattering on the positions and widths
of CIRs under isotropic transversal confinement has been investigated~\cite{Melezhik2011}.

In this work, we study the width of the $s$-wave CIR in quasi-1D gases under tunable transversely anisotropic confinement using the Fermi-Huang pseudopotential approach~\cite{Olshanii1998}.
We propose to tune the width of the CIR with the transverse frequency ratio.
Furthermore, we carefully study the thermodynamics of this system and find the change of the CIR width can manifest itself in the position of the discontinuity in the interaction energy density.
At first, we analytically calculate the resonance position $B_{1D}$ and width $\Delta_{1D}$ of the CIR.
We find that both $B_{1D}$ and $\Delta_{1D}$ can be tuned by varying the ratio between the two transverse trapping frequencies.
Typically, we find that, for quasi-1D $^{133}$Cs Bose gases, $\Delta_{1D}$ diverges at two critical frequency ratios $\eta_{c1}$ and $\eta_{c2}$, while there is no divergence in the CIR width for $^{40}$K atoms.
Furthermore, we study the normal-state thermodynamics of the quasi-1D atomic gases across the CIRs using the quantum virial expansions.
Interestingly, the scattering branch of the interaction energy shows an abrupt discontinuity at the magnetic field $B=B_{1D}+\Delta_{1D}$,
which can be detected in experiment.

This paper is organized as follows. In Sec.~\ref{2}, we present the model Hamiltonian.
We derive the resonance width and position of the $s$-wave CIRs as functions of the transverse anisotropy.
In Sec.~\ref{3}, we discuss the corresponding normal-state thermodynamics of the quasi-1D atomic gases using the quantum virial expansions.
Finally, we give a summary in Sec.~\ref{4}.

\section{CIR with transverse anisotropy}\label{2}

\subsection{General formalism}\label{2.1}

We consider two interacting atoms in a harmonic trap with tight confinements in the transverse $x-y$ plane and a weak confinement in the axial $z$-direction. The transverse confinements are anisotropic, i.e., $\omega_{x}=\eta\omega_{y}\gg\omega_{z}$ with transverse trapping frequency ratio $\eta=\omega_{x}/\omega_{y}$, where $\omega_{\xi} (\xi=x,y,z)$ is the trapping frequency in $\xi$ axis.
Near an $s$-wave Feshbach resonance, the interaction between the atoms is dominated by the $s$-wave scattering, and the Hamiltonian for the relative atomic motion can be written as
\begin{equation}\label{3D relative Hamiltonian}
\hat{H}_{rel}=-\frac{\hbar^{2}}{2m_{r}}\frac{\partial^{2}}{\partial z^{2}} + \hat{H}_{\perp} + g_{3D}\delta^{(3)}(\vec{r})\frac{\partial}{\partial r}(r\cdot),
\end{equation} where
\begin{equation}\label{3D x-y Hamiltonian}
\hat{H}_{\perp}=-\frac{\hbar^{2}}{2m_{r}}\left(\frac{\partial^{2}}{\partial x^{2}}+\frac{\partial^{2}}{\partial y^{2}}\right)+\frac{1}{2}m_{r}\omega_{y}^{2}\left(\eta^{2}x^{2}+y^{2}\right).
\end{equation}
Here, $g_{3D}=4\pi\hbar^{2}a_{3D}/M$ is the three-dimensional coupling rate, $\hbar$ is the reduced Planck constant, $a_{3D}$ is the three-dimensional $s$-wave scattering length,
and $m_{r}=M/2$ is the reduced mass for the relative motion with atom mass $M$.

Following the standard renormalization procedure~\cite{Cui2012,renormalization2002,renormalization2005}, we write the energy-dependent (or $k$-dependent) three-dimensional $s$-wave scattering length $a_{3D}(k)$ as
\begin{eqnarray}\label{a3Dk}
a_{3D}(k)=a_{bg}\left[1+\frac{\Delta}{E(k)/\delta\mu-(B-B_{0})}\right],
\end{eqnarray}
where $E(k)=\hbar^2k^2/M+(\eta+1)\hbar\omega_{y}/2$, $a_{bg}$ is the background scattering length, $B$ is the magnetic-field strength,
$B_0$ is the position of the Feshbach resonance, $\Delta$ is the three-dimensional resonance width,
and $\delta\mu$ is the magnetic-moment difference between the atom state and the closed-channel molecular state.
The expression of energy-dependent scattering length in Eq.~(\ref{a3Dk}) is applicable for both wide and narrow Feshbach resonances.

The quasi-1D scattering amplitude can be written as~\cite{review2015,review2008,ps1990}
\begin{eqnarray}\label{amplitude}
f_{e}(k) = \frac{-1}{1+i\cot\delta_{s}(k)}.
\end{eqnarray}
Here, the phase shift $\delta_{s}(k)$ is given by
\begin{eqnarray}\label{phase shift}
\cot\delta_{s}(k)=-\frac{kd^{2}}{2\sqrt{\eta}a_{3D}}\left[1-C(\eta)\frac{a_{3D}}{d}\right],
\end{eqnarray}
where $d=\sqrt{2\hbar/\left(M\omega_{y}\right)}$ and
\begin{eqnarray}\label{C}
C(\eta)&&=\frac{1}{\sqrt{\pi}}\int\limits _{0}^{\infty}du\left\{ \frac{1}{u^{3/2}}- \sqrt{\frac{\eta}{u}} \right. \nonumber\\
&&\left. ~~\times\left[\frac{1}{\sqrt{\left(1-e^{-\eta u}\right)\left(1-e^{-u}\right)}}-1\right]\right\}.
\end{eqnarray}

The effective 1D coupling rate is given by $g_{1D}=\lim_{k\rightarrow0}\frac{\hbar^{2}k}{m_{r}}\frac{\mathop{\rm Re}f_{e}(k)}{\mathop{\rm Im}f_{e}(k)}$~\cite{Olshanii1998,Olshanii2003,Peng2010}.
Using Eqs.~(\ref{a3Dk}) and (\ref{amplitude}), we derive
\begin{eqnarray}\label{g1Dk}
g_{1D}(k)=g_{bg}\left[1+\frac{\Delta_{1D}}{\hbar^2k^2/(M\delta\mu)-(B-B_{1D})}\right],
\end{eqnarray}
where $g_{bg}=2\sqrt{\eta}\gamma\hbar\omega_{y} a_{bg}$ is the background effective 1D coupling strength,
\begin{eqnarray}\label{B1D}
B_{1D}=B_{0}-\left[\frac{1}{1-C(\eta)a_{bg}/d}-1\right]\Delta+\frac{(\eta+1)\hbar\omega_{y}}{2\delta\mu},
\end{eqnarray} and
\begin{eqnarray}\label{Delta1D}
\Delta_{1D}=\frac{\Delta}{1-C(\eta)a_{bg}/d}.
\end{eqnarray}
Here, $|\Delta_{1D}| \gg (\eta + 1)\hbar\omega_{y}/\delta\mu$ corresponds to the wide CIR limit,
and $|\Delta_{1D}| \ll (\eta + 1)\hbar\omega_{y}/\delta\mu$ corresponds to the narrow CIR limit~\cite{Cui2012}.

Near CIR ($B\sim B_{1D}$) and for
$\hbar^2k^2/M\ll |\Delta_{1D}\delta\mu|$, one has
\begin{equation}\label{g1Dkeff}
\frac{1}{g_{1D}(k)}\simeq\frac{1}{g_{1D}}- \frac{1}{2}r_{1D}k^2,
\end{equation}
where
\begin{eqnarray}\label{r1D}
r_{1D}=-\frac{2\hbar^2}{Mg_{bg}\Delta_{1D}\delta\mu}
\end{eqnarray} is the 1D effective range characterizing
$k$-dependence in $g_{1D}(k)$, and the zero-energy coupling strength $g_{1D}$ is
\begin{eqnarray}\label{g1Dk0}
g_{1D}=g_{bg}\left(1-\frac{\Delta_{1D}}{B-B_{1D}}\right),
\end{eqnarray} which diverges at $B=B_{1D}$.
Therefore, the $B_{1D}$ is called the resonance position of a CIR, and $\Delta_{1D}$ is called the resonance width of a CIR.
We will show in the next subsection that the width of the CIR can be tuned from wide to narrow by varying $\eta$.

\subsection{Numerical calculations}\label{2.2}

\begin{figure}[ht]
\includegraphics[width=9cm]{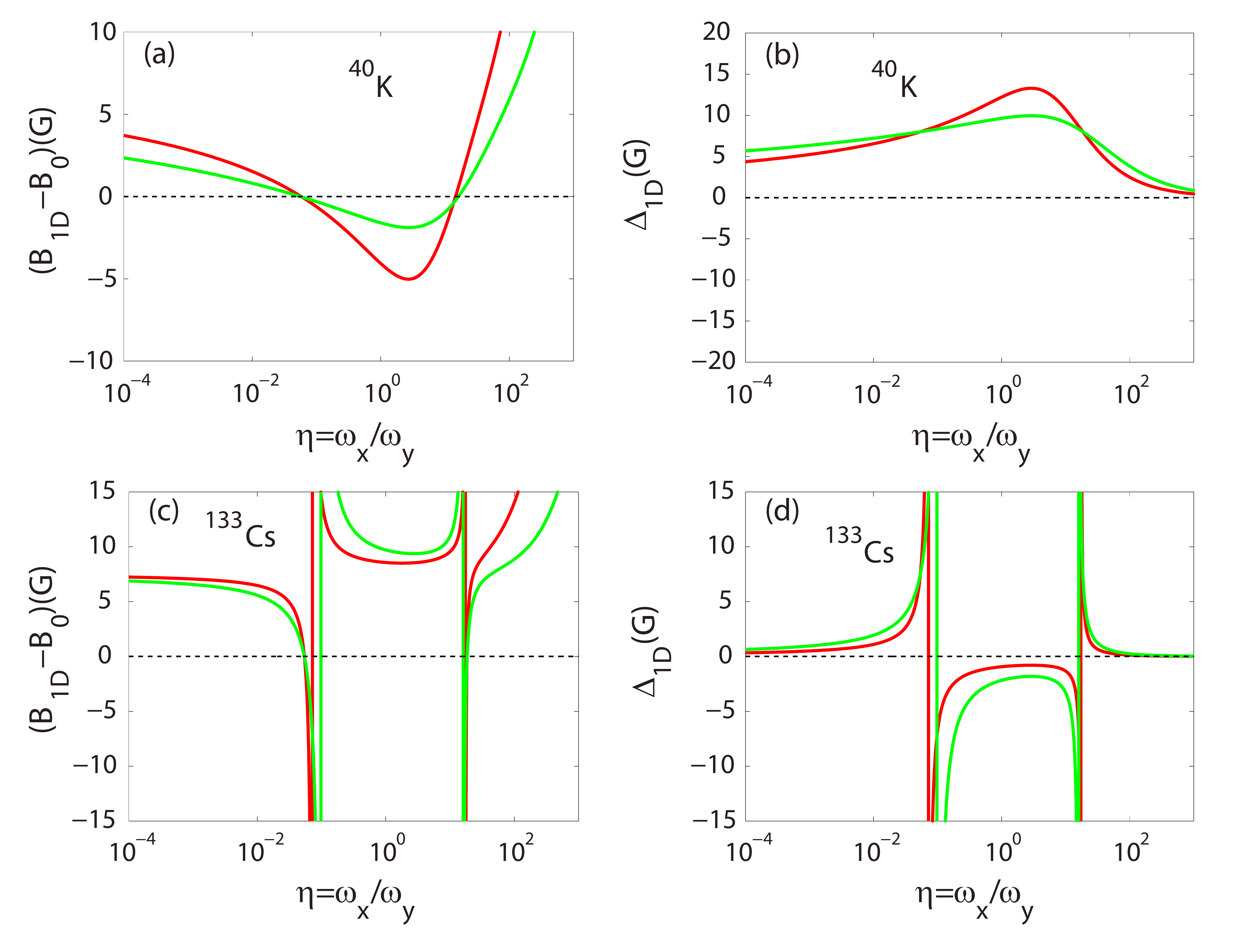}
\caption{(Color online).
(a) and (b) ( (c) and (d) ) show the resonance position $B_{1D}$ and width $\Delta_{1D}$ of $^{40}$K ( $^{133}$Cs ) atomic gas as functions of $\eta$, respectively.
Here, we choose $\omega_{y}=(2\pi)\times330$kHz (dark red curves) and $\omega_{y}=(2\pi)\times80$kHz (light green curves) for comparison.
For $^{40}$K fermions, the Feshbach resonance is at $B_{0}=202.1$G, and the three-dimensional width is $\Delta=8$G~\cite{review2010,K40s}.
For $^{133}$Cs bosons, $B_{0}=547$G and $\Delta=7.5$G~\cite{review2010,Cs1332004,Cs1332009}.}\label{B1Dfig}
\end{figure}

In Fig.~\ref{B1Dfig}, we plot the CIR position $B_{1D}$ and width $\Delta_{1D}$ as functions of $\eta$ for $^{40}$K and $^{133}$Cs atoms, respectively.

For $^{133}$Cs atoms with $\omega_{y}=(2\pi)\times330$kHz, the CIR width $\Delta_{1D}$ diverges at $\eta_{c1}\simeq0.07$ and $\eta_{c2}\simeq17.29$, near which the CIR is wide.
However, as $\eta$ changes, the condition for a wide CIR may not hold.
For example, with $\eta=21$, $|\Delta_{1D}| \simeq 2.55G$, which is on the same order of $(\eta + 1)\hbar\omega_{y}/\delta\mu \simeq 2.90G$.

The divergent or nonmonotonic behavior of $^{133}$Cs is connected to the transformation from the quasi-1D geometry to the quasi-2D geometry by continuously changing the transverse anisotropy~\cite{Peng2010,Zhang2011}. Therefore, the divergent point is the point where the quasi-1D system goes into the quasi-2D system.

As shown in Fig.~\ref{B1Dfig}, the width and resonance position of the CIR for $^{133}$Cs diverge at the same critical value of $\eta_c$. Physically, the divergences of the width and resonance position mean that $^{133}$Cs has no resonance position at the critical $\eta_{c}$.

On the other hand, while there is no divergence in the CIR width for $^{40}$K atoms, the condition for the wide CIR typically holds, except for unrealistically large $\eta$'s. As we will show in the following sections, these changes in the character of the CIR would manifest themselves in the normal-state many-body properties of the quasi-1D atomic gases, when the transverse anisotropy is tuned.

The divergence in the width of the CIR is system dependent. Comparing $^{40}$K with $^{133}$Cs as shown in Fig.~\ref{B1Dfig}, it is found that the divergence in the widths of the CIR is system dependent. The condition for this divergence is \begin{eqnarray}\label{condition}
C(\eta)=\frac{1}{a_{bg}}\sqrt{\frac{2\hbar}{M\omega_{y}}},
\end{eqnarray} i.e., the denominator of Eq.~(\ref{Delta1D}) equals zero.
The atomic system with specific value of $a_{bg}$ and atom mass $M$ which satisfy Eq.~(\ref{condition}) can have divergent width, otherwise there is no divergence in the width.
For $^{133}$Cs, $a_{bg}=2500a_{0}$ which is easy to satisfy the divergent condition. However, for $^{40}$K, $a_{bg}=174a_{0}$ which is much smaller than the one of $^{133}$Cs and it cannot satisfy the divergent condition.
Therefore, the system with large $a_{bg}$ can reach the divergent condition more easily.

\section{Thermodynamics in normal state}\label{3}

In this section, we study the normal-state properties of quasi-1D atomic gases using the quantum virial expansions.
For consistency, we assume that the temperature is high enough for the gases to be in the normal state, but not too high to make the excited states in the transverse directions are not thermally populated, i.e., $T \ll (\eta+1)\hbar\omega_{y}/k_{B}$, where $T$ is the system temperature and $k_B$ is the Boltzmann constant.
Here, we mainly focus on the behavior of the interaction energy density, which can be measured in experiment.

\subsection{Fermions}\label{3.1}

Based on the local density approximation, the thermodynamic potential of a non-interacting two-component Fermi gas with
equal spin populations in our quasi-1D geometry takes the form
\begin{eqnarray}\label{omega1}
\Omega_{ideal}(z) && = -\frac{2}{\beta}\int_{-\infty}^{\infty}dx\int_{-\infty}^{\infty}dy\int_{-L/2}^{L/2}dz\int\frac{d^{3}\vec{k}}{(2\pi)^3}\nonumber\\
&&~~\times\ln\left[1+ze^{-\beta\left(\frac{\hbar^{2}k^{2}}{2M}+\frac{M\omega_{x}^{2}x^{2}}{2}+\frac{M\omega_{y}^{2}y^{2}}{2}\right)}\right]\nonumber\\
&& = -\frac{8\sqrt{2}}{\eta}\frac{L}{\lambda}\frac{(k_{B}T)^3}{(\hbar\omega_y)^2}f_{7/2}(z),
\end{eqnarray}
where $L$ is the system size in the axial direction, $z \equiv \text{exp}( \beta\mu )$ is the fugacity,
$\mu$ is the corresponding chemical potential, $\beta=1/(k_B T)$, $\lambda\equiv\sqrt{2\pi\hbar^{2}/(Mk_{B}T)}$ is the thermal de Broglie wavelength, and $f_{\upsilon}(z)=[1/\Gamma(\upsilon)]\int_{0}^{\infty}[x^{\upsilon-1}/(z^{-1}e^{x}+1)]dx$ is the standard Fermi-Dirac integral
with the gamma function $\Gamma(\upsilon)$~\cite{Pathria1996}.

Accordingly, we can rewrite the thermodynamic potential of a strongly interacting Fermi gas as (up to the second order)~\cite{liureview2013,qin2016,PengPLA2011,PengPRA2011}
\begin{eqnarray}\label{omegah}
\Omega
=-\frac{8\sqrt{2}}{\eta}\frac{L}{\lambda}\frac{(k_{B}T)^3}{(\hbar\omega_y)^2}\left[f_{7/2}(z) + z^{2}\Delta b_{2} \right],
\end{eqnarray} where $\Delta b_{2}$ is the second virial coefficient.
Therefore, the particle number density of atoms is given by
\begin{eqnarray}\label{nh}
n&&=-\frac{1}{L}\frac{\partial\Omega}{\partial\mu}\nonumber\\
&& = \frac{8\sqrt{2}}{\eta}\frac{1}{\lambda}\left(\frac{k_{B}T}{\hbar\omega_y}\right)^2\left[ f_{5/2}(z) + 2z^{2}\Delta b_{2}\right]\nonumber\\
&& \simeq \frac{8\sqrt{2}}{\eta}\frac{1}{\lambda}\left(\frac{k_{B}T}{\hbar\omega_y}\right)^2\left(z-2^{-5/2}z^{2}+ 2z^{2}\Delta b_{2}\right),
\end{eqnarray}
where we use~\cite{Pathria1996}
\begin{eqnarray}\label{fH}
f_{\upsilon}(z)=z-\frac{z^{2}}{2^{\upsilon}}+\frac{z^{3}}{3^{\upsilon}}-\cdot\cdot\cdot
\end{eqnarray} at high temperatures.
Then we can derive the specific expression for fugacity:
\begin{eqnarray}\label{zh}
z
&& \simeq \frac{\eta}{8\sqrt{2}}\left(\frac{\hbar\omega_y}{k_{B}T}\right)^2n\lambda +2^{-5/2}\left[\frac{\eta}{8\sqrt{2}}\left(\frac{\hbar\omega_y}{k_{B}T}\right)^2n\lambda\right]^{2} \nonumber\\
&&~~- 2\left[\frac{\eta}{8\sqrt{2}}\left(\frac{\hbar\omega_y}{k_{B}T}\right)^2n\lambda\right]^{2}\Delta b_{2}.
\end{eqnarray}

The logarithm of the grand canonical partition function $\Xi$ can be represented by
\begin{eqnarray}\label{Xih}
\ln\Xi&&=-\frac{\Omega}{k_{B}T}\nonumber\\
&&=\frac{8\sqrt{2}}{\eta}\frac{L}{\lambda}\left(\frac{k_{B}T}{\hbar\omega_y}\right)^2\left[f_{7/2}(z) + z^{2}\Delta b_{2}\right].
\end{eqnarray}
Based on the above equations, we derive the internal energy density
\begin{eqnarray}\label{Eh}
\epsilon&&\equiv-\frac{1}{L}\left(\frac{\partial\ln\Xi}{\partial\beta}\right)_{z}\nonumber\\
&&=\frac{8\sqrt{2}}{\eta}\frac{k_{B}T}{\lambda}\left(\frac{k_{B}T}{\hbar\omega_y}\right)^2\frac{5}{2}\left[f_{7/2}(z) + z^{2}\left(\Delta b_{2}+\frac{2}{5}T\Delta b^{'}_{2}\right)\right] \nonumber\\
&&\simeq\frac{5}{2}\frac{k_{B}T}{\lambda}\left[n\lambda + 2^{-7/2}\frac{\eta}{8\sqrt{2}}\left(\frac{\hbar\omega_y}{k_{B}T}\right)^2(n\lambda)^{2} \right. \nonumber\\
&& \left.~~ +\frac{\eta}{8\sqrt{2}}\left(\frac{\hbar\omega_y}{k_{B}T}\right)^2(n\lambda)^{2}\left(-\Delta b_{2}+\frac{2}{5}T\Delta b^{'}_{2}\right)\right],
\end{eqnarray} where $\Delta b^{'}_{2}=d(\Delta b_{2})/dT$.

Finally, we can obtain the dimensionless interaction energy density~\cite{Ho2004s,Ho2004p}:
\begin{eqnarray}\label{eint1D}
\epsilon_{int}=-\Delta b_{2}+\frac{2}{5}T\Delta b^{'}_{2}.
\end{eqnarray}

\subsection{Bosons}\label{3.2}

Following the similar derivations, the internal energy density for the quasi-1D bosons is given by
\begin{eqnarray}\label{EhB}
\epsilon
&&\simeq \frac{5}{2}\frac{k_{B}T}{\lambda}\left[n\lambda - 2^{-7/2}\frac{\eta}{4\sqrt{2}}\left(\frac{\hbar\omega_y}{k_{B}T}\right)^2(n\lambda)^{2}\right. \nonumber\\
&& \left.~~ + \frac{\eta}{4\sqrt{2}}\left(\frac{\hbar\omega_y}{k_{B}T}\right)^2(n\lambda)^{2}\left(-\Delta b_{2}+\frac{2}{5}T\Delta b^{'}_{2}\right)\right].
\end{eqnarray}
Therefore, we have the dimensionless interaction energy density:
\begin{eqnarray}\label{eint1DB}
\epsilon_{int}=-\Delta b_{2}+\frac{2}{5}T\Delta b^{'}_{2}.
\end{eqnarray}

\subsection{Fugacity}\label{3.3}

\begin{figure}[ht]
\includegraphics[width=9cm]{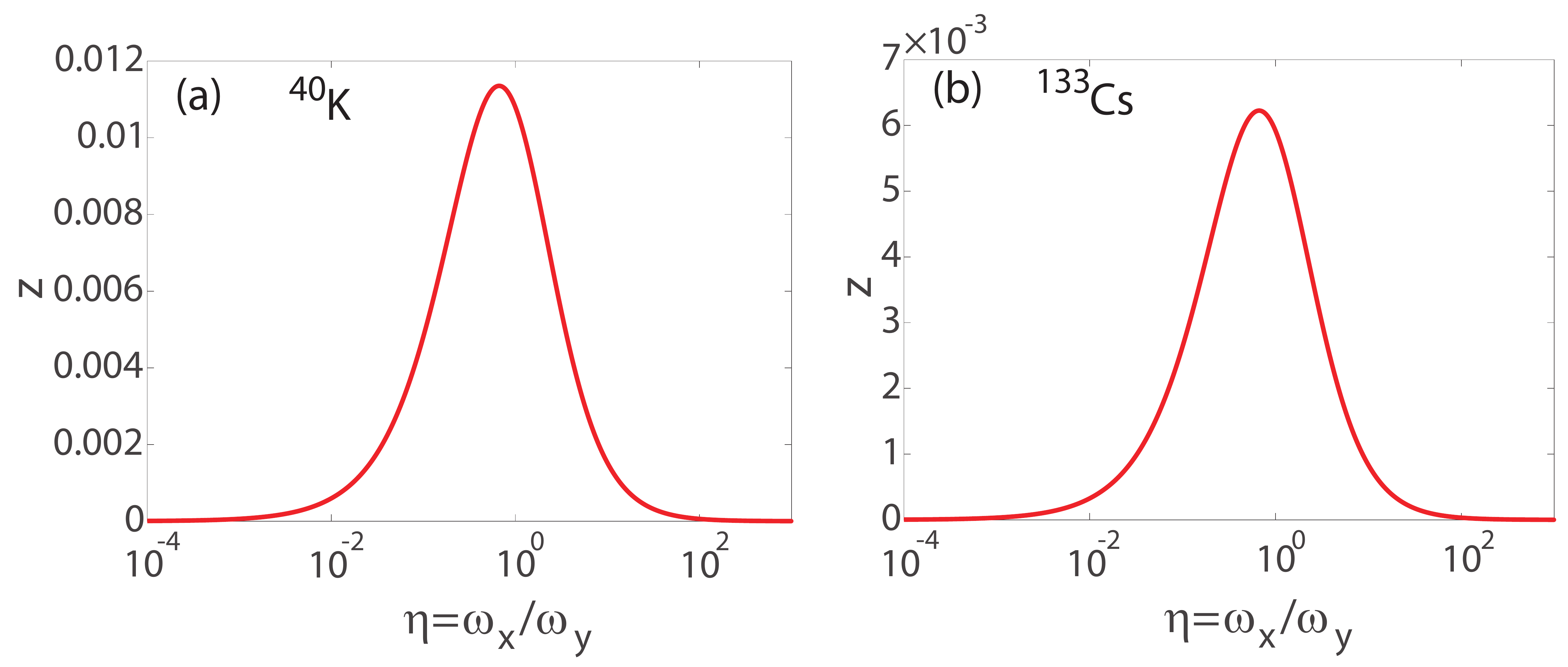}
\caption{(Color online).
(a) and (b) show the fugacities $z$ as functions of $\eta$ at $T=(\eta+1)\hbar\omega_y/k_B$ for $^{40}$K fermions and $^{133}$Cs bosons, respectively.
Here, we choose $\omega_{y}=(2\pi)\times330$kHz.}\label{z1Dfig}
\end{figure}

In order to justify the condition of the virial expansion, we calculate the fugacity at $T=(\eta+1)\hbar\omega_y/k_B$ as below.

Substituting $T=(\eta+1)\hbar\omega_y/k_B$ into Eq.~(\ref{zh}), we have
\begin{eqnarray}\label{zh2}
z
&& =\eta\left(\frac{1}{1+\eta}\right)^{5/2}\frac{n}{8\sqrt{2}}\sqrt{\frac{2\pi\hbar}{M\omega_y}} \nonumber\\
&&~~ + 2^{-5/2}\left[\eta\left(\frac{1}{1+\eta}\right)^{5/2}\frac{n}{8\sqrt{2}}\sqrt{\frac{2\pi\hbar}{M\omega_y}}\right]^{2} \nonumber\\
&&~~ - 2\left[\eta\left(\frac{1}{1+\eta}\right)^{5/2}\frac{n}{8\sqrt{2}}\sqrt{\frac{2\pi\hbar}{M\omega_y}}\right]^{2}\Delta b_{2}.
\end{eqnarray}

For the typical experimental parameter, the particle number density is $n\simeq10^7$m$^{-1}$~\cite{density20041,density20042}.
Therefore, with $\omega_{y}=(2\pi)\times330$kHz, we have $n\sqrt{2\pi\hbar/(M\omega_y)}/(8\sqrt{2})\simeq0.06\ll1$ for $^{40}$K fermions, and $n\sqrt{2\pi\hbar/(M\omega_y)}/(8\sqrt{2})\simeq0.03\ll1$ for $^{133}$Cs bosons.

Keeping up to the $n\sqrt{2\pi\hbar/(M\omega_y)}/(8\sqrt{2})$ term of Eq.~(\ref{zh2}), we have
\begin{eqnarray}\label{zh3}
z \simeq \eta\left(\frac{1}{1+\eta}\right)^{5/2}\frac{n}{8\sqrt{2}}\sqrt{\frac{2\pi\hbar}{M\omega_y}}.
\end{eqnarray}

As shown in Fig.~\ref{z1Dfig}, we calculate the fugacity as a function of $\eta$ at $T=(\eta+1)\hbar\omega_y/k_B$ numerically.
It is found that the fugacities always satisfy $z\ll1$ for both $^{40}$K and $^{133}$Cs atoms, i.e., the virial expansion might be used even when the temperature is much lower than the zero-point energy.

\subsection{Second virial coefficients}\label{3.4}

To calculate the interaction energies in Eqs.~(\ref{eint1D},~\ref{eint1DB}), we need to evaluate the second virial coefficients $\Delta b_2$ at first.
They can be expressed in terms of the phase shifts of the corresponding two-body scattering problem.
The second virial coefficient $\Delta b_{2}$ for the repulsive scattering branch takes the form~\cite{Ho2004s,Ho2004p}
\begin{eqnarray}\label{b2sc}
\Delta b_{2}^{sc}
&&=\int_{0}^{\infty}e^{-\hbar^2k^{2}/(Mk_{B}T)}\frac{\partial\delta_{s}(k)}{\partial k}\frac{dk}{\pi}\nonumber\\
&&=\frac{d}{2\pi\sqrt{\eta}}\int_{0}^{\infty}\frac{[d/a_{3D}-C(\eta)]e^{-\hbar^2k^{2}/(Mk_{B}T)}}{1+\frac{k^{2}d^{2}}{4\eta}\left[d/a_{3D}-C(\eta)\right]^{2}}dk.\nonumber\\
\end{eqnarray}
For the attractive branch, $\Delta b_{2}$ is given by
\begin{align}\label{b2bd}
\Delta b_{2}^{bd}
&=e^{|E_{b}|/(k_{B}T)}\nonumber\\
&~~+ \frac{d}{2\pi\sqrt{\eta}}\int_{0}^{\infty}\frac{[d/a_{3D}-C(\eta)]e^{-\hbar^2k^{2}/(Mk_{B}T)}}{1+\frac{k^{2}d^{2}}{4\eta}\left[d/a_{3D}-C(\eta)\right]^{2}}dk.
\end{align}

For wide CIR, the binding energy $E_b$ ($E_b<0$) for the quasi-1D system is determined by~\cite{Peng2010,Zhang2011,review2008,Taylor1972}
\begin{eqnarray}\label{Eb-wide}
\frac{d}{a_{3D}}=\frac{1}{\sqrt{\pi}}\int\limits _{0}^{\infty}du\left[ \frac{1}{u^{3/2}} - \frac{\sqrt{\eta}e^{uE_{b}/(2\hbar\omega_{y})}}{\sqrt{u\left(1-e^{-\eta u}\right)\left(1-e^{-u}\right)}} \right],\nonumber\\
\end{eqnarray}
where the corresponding $a_{3D}$ is $k$-independent, i.e., $a_{3D}=a_{bg}[1-\Delta/(B-B_{0})]$.

When the condition for the wide CIR is not satisfied, the binding energy of a shallow bound state can be obtained from
\begin{eqnarray}\label{Eb-narrow}
\frac{d}{a_{3D}(E_{b})}=\frac{1}{\sqrt{\pi}}\int\limits _{0}^{\infty}du\left[ \frac{1}{u^{3/2}} - \frac{\sqrt{\eta}e^{uE_{b}/(2\hbar\omega_{y})}}{\sqrt{u\left(1-e^{-\eta u}\right)\left(1-e^{-u}\right)}} \right],\nonumber\\
\end{eqnarray} where the $E_{b}$-dependent $a_{3D}(E_{b})$ is
\begin{eqnarray}\label{a3DEb}
a_{3D}(E_{b})=a_{bg}\left\{1 - \frac{\Delta}{B - \left[B_{0} + \frac{(\eta+1)\hbar\omega_{y}}{2\delta\mu} + \frac{E_{b}}{\delta\mu}\right]}\right\}.\nonumber\\
\end{eqnarray}

\subsection{Numerical calculations}\label{3.5}

Following the above analytical expressions, we numerically calculate the second virial coefficients and the interaction energy densities using the typical experimental parameters for $^{40}$K and $^{133}$Cs atoms, respectively.

\begin{figure}[ht]
\includegraphics[width=9cm]{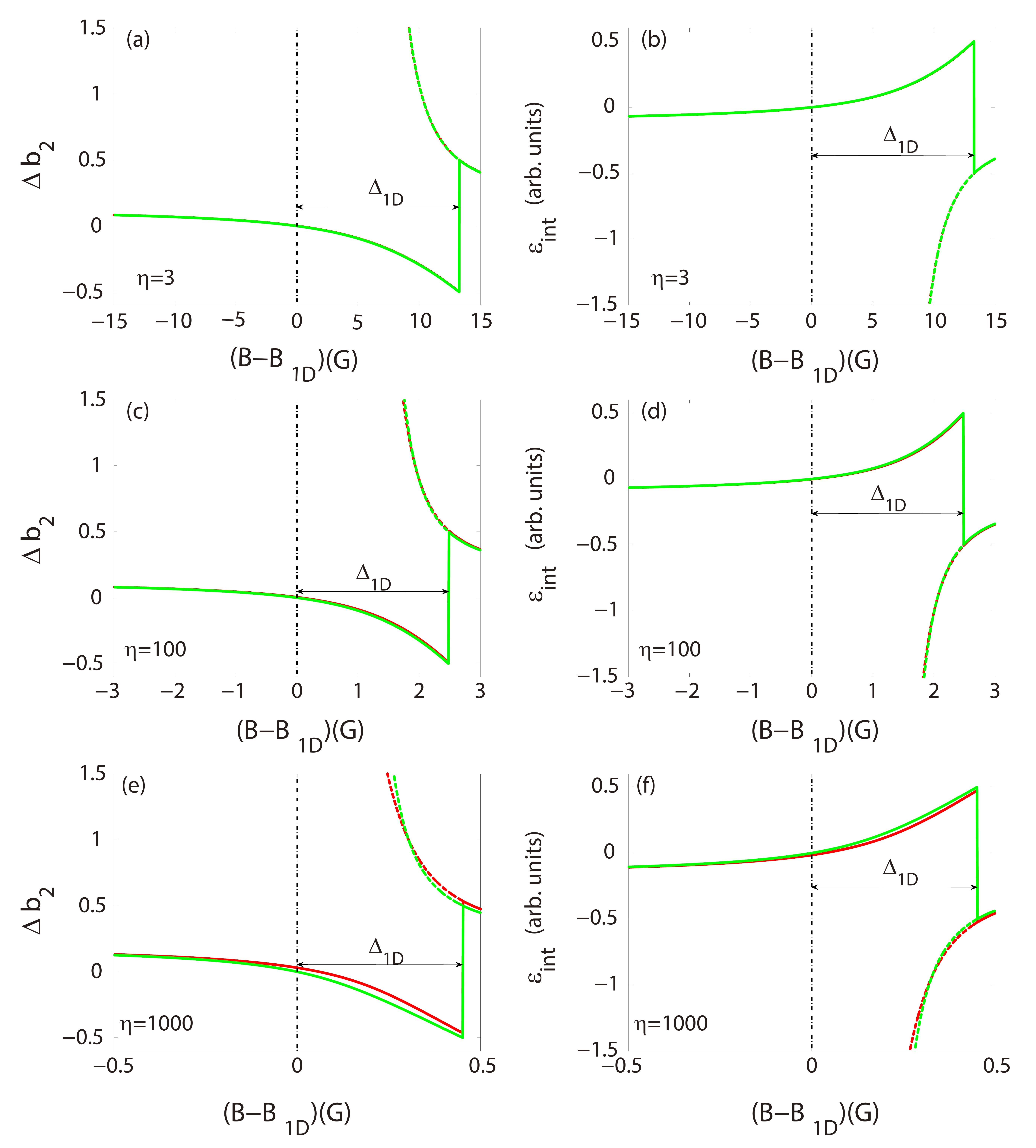}
\caption{(Color online). Second virial coefficient of
two-species $^{40}$K fermions across the CIRs at $T=5\mu K$ for (a) $\eta=3$, (c) $\eta=100$, and (e) $\eta=1000$.
Interaction energy for (b) $\eta=3$, (d) $\eta=100$, and (f) $\eta=1000$.
Here, we choose $k$-dependent (dark red curves) and $k$-independent (light green curves) for comparison with fixed $\omega_{y}=(2\pi)\times330$kHz.
The Feshbach resonance is at $B_{0}=202.1$G with width $\Delta=8$G, $a_{bg}\simeq 174a_{0}$ and $\delta\mu=1.68\mu_{B}$ for $^{40}$K fermions, where $a_{0}\simeq0.529\times10^{-10}$m is the Bohr radius
and $\mu_{B}\simeq 9.274\times10^{-28}$J/G is the Bohr magneton~\cite{review2010,K40s}.
Solid and dashed lines are respectively for scattering and attractive branches.}\label{Kfig}
\end{figure}

\begin{figure}[ht]
\includegraphics[width=9cm]{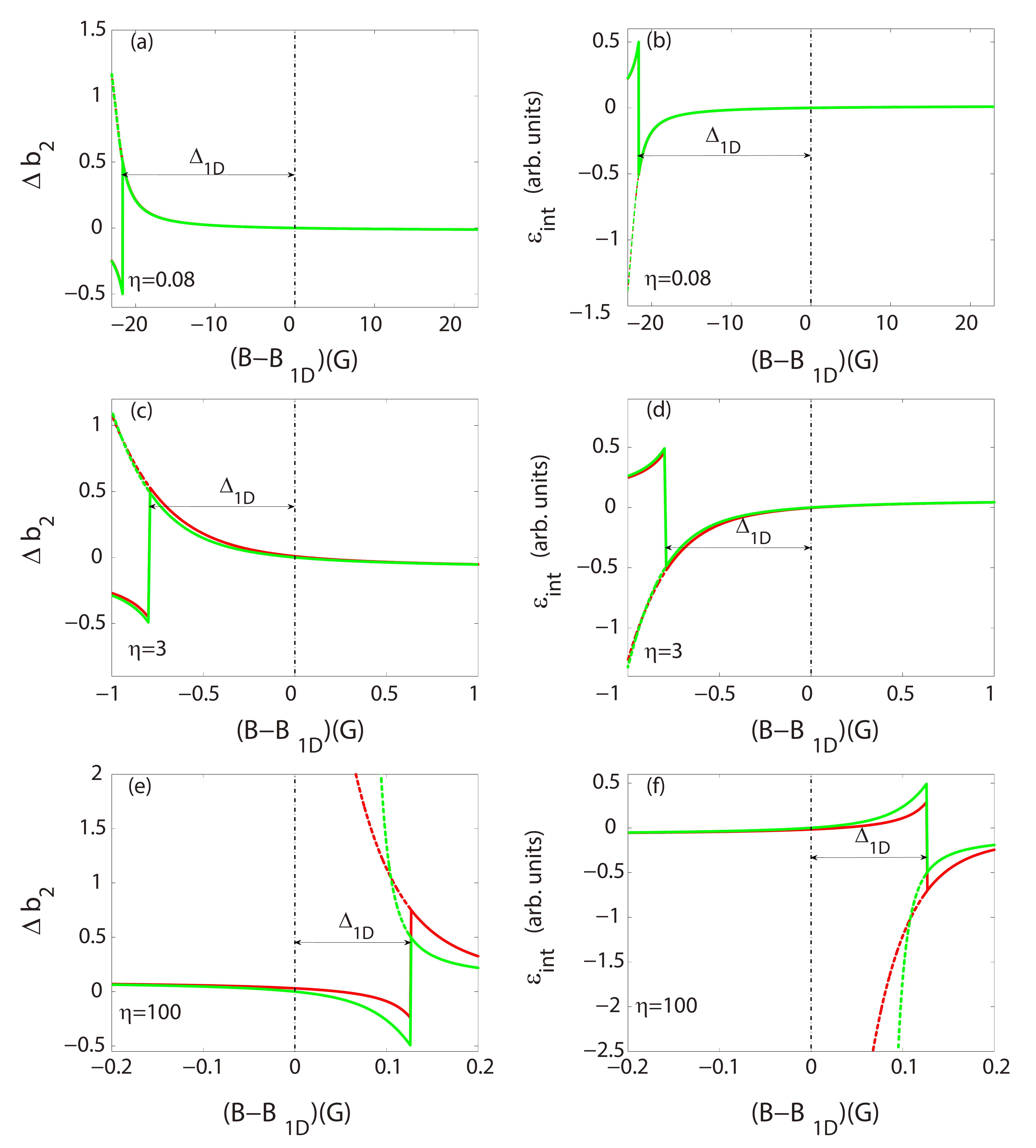}
\caption{(Color online). Second virial coefficient of
two-species $^{133}$Cs bosons across the CIRs at $T=5\mu K$ for (a) $\eta=0.08$, (c) $\eta=3$, and (e) $\eta=100$.
Interaction energy for (b) $\eta=0.08$, (d) $\eta=3$ and (f) $\eta=100$.
Here, we choose $k$-dependent (dark red curves) and $k$-independent (light green curves) for comparison with fixed $\omega_{y}=(2\pi)\times330$kHz.
The Feshbach resonance is at $B_{0}=547$G with width $\Delta=7.5$G, $a_{bg}\simeq 2500a_{0}$, and $\delta\mu=1.79\mu_{B}$ for $^{133}$Cs bosons~\cite{review2010,Cs1332004,Cs1332009}.
Solid and dashed lines are respectively for scattering and attractive branches.}\label{Csfig}
\end{figure}

In Fig.~\ref{Kfig}, we plot $\Delta b_{2}$ and $\epsilon_{int}$ for two-species $^{40}$K fermions across CIR with different $\eta$.
The solid and dashed lines are respectively for scattering and attractive branches shown in Figs.~\ref{Kfig} and \ref{Csfig}.
The dashed line denotes the molecular state which includes the two-body binding energy in the interaction energy and second virial coefficient, while the solid line dose not include the binding energy.
It shows that the scattering branch goes through an abrupt discontinuity, the position of which is given by $B=B_{1D}+\Delta_{1D}$~\cite{Cui2012}.
Physically, the discontinuity point in the scattering branch of the interaction energy density is where the binding energy goes beyond its threshold for the existence of the molecular state, i.e., the two-body bound state transforms to a scattering state leads to the discontinuity in the scattering branch of the interaction energy density~\cite{narrow2012}.

From $\eta=3$ to $\eta=1000$, we find that $\Delta b_{2}$ and $\epsilon_{int}$ exhibit different features along a wide CIR to a narrow CIR.
For example, from a wide CIR to a narrow CIR, the change of the CIR width dramatically shifts the position of the discontinuity point of $\Delta b_{2}$ and $\epsilon_{int}$ curves.
Therefore, the tunability of the CIR width can leave detectable signatures in the normal-state properties of the many-body system.

In Fig.~\ref{Csfig}, we show $\Delta b_{2}$ and $\epsilon_{int}$ for $^{133}$Cs atoms across a CIR with different $\eta$. While the general conclusions are similar to the cases of $^{40}$K, the tunability of the CIR is even more obvious.

Why there are different features in the interaction energy densities for the wide and narrow CIRs?
The wide (narrow) CIR corresponds to $|r_{1D}|\ll|g_{bg}|$ ($|r_{1D}|\gg|g_{bg}|$)~\cite{narrow2012}.
From Eq.~\eqref{r1D}, it is found that the resonance width $\Delta_{1D}$ is proportional to $1/|r_{1D}|$, so that $|r_{1D}|$ is a very important interaction parameter to describe the wide and narrow CIRs. Since $|r_{1D}|\gg|g_{bg}|$, the two-body effective 1D coupling strength for the narrow CIR is $k$-dependent (Eq.~\eqref{g1Dkeff}), i.e., it must be described by two interaction parameters: $k$-independent effective 1D coupling strength $g_{1D}$ (Eq.~\eqref{g1Dk0}) and $r_{1D}$, while the effective 1D coupling strength for the wide CIR is $k$-independent (Eq.~\eqref{g1Dk0}), because the effective range is too small compared to $|g_{bg}|$ across a wide resonance: $|r_{1D}|\ll|g_{bg}|$~\cite{narrow2012}. Therefore, the two-body effective 1D coupling strengths for the narrow and wide CIRs have completely different features. Furthermore, the interaction energy is directly connected to the two-body effective 1D coupling strength or the interaction parameters.
Therefore, the interaction energy across a narrow CIR must be described by two interaction parameters: $k$-independent $g_{1D}$ and $r_{1D}$, while the interaction energy across a wide CIR can be described by only one interaction parameter $g_{1D}$, because the effective range compared to $|g_{bg}|$ is too small to contribute to the interaction energy across a wide resonance~\cite{narrow2012,narrow2014}.

Futhermore, we compare the $k$-dependent (or energy-dependent) curve with the corresponding $k$-independent curve in Figs.~\ref{Kfig} and \ref{Csfig}.
It is found that the two curves almost coincide with each other for a wide CIR.
However, for a narrow CIR, the two curves deviate from each other.
Therefore, the $k$-dependent result is more accurate for a narrow CIR than the $k$-independent one, while for a wide CIR, both of them are valid.

\subsection{Addition}\label{3.6}

Refs.~\cite{PengPLA2011,PengPRA2011} give another way to calculate the normal-state thermodynamics for the interacting Fermi gases in a three-dimensional anisotropic trap.
They start from the Hamiltonian for the relative atomic motion~\cite{PengPLA2011,PengPRA2011}
\begin{equation}\label{3D relative Hamiltonian2}
\hat{H}_{rel}=-\frac{\hbar^{2}}{2m_{r}}\nabla_{\vec{r}}^{2} + \frac{1}{2}m_{r}\omega_{z}^{2}\left(\tilde{\eta}^{2}\rho^{2}+z^{2}\right) + g_{3D}\delta^{(3)}(\vec{r})\frac{\partial}{\partial r}(r\cdot),
\end{equation} where $\vec{r}=(z,\rho)$ is the relative coordinate, $\vec{\rho}=(x,y)$, $\tilde{\eta}=\omega_{\perp}/\omega_{z}$, and $\omega_{\perp}=\omega_{x}=\omega_{y}$.
Different from Eq.~(\ref{3D relative Hamiltonian2}), in our model (Eq.~(\ref{3D relative Hamiltonian})), we ignore the confinement in the axial $z$-direction, i.e., $\omega_{z}=0$, and the particle can move freely in the axial $z$-direction.

By solving the corresponding Schr{\" o}dinger equation, one can obtain the second virial coefficient~\cite{PengPLA2011,PengPRA2011}.
Furthermore, the corresponding normal-state thermodynamic potential for interacting Fermi gases can be calculated.

From the thermodynamic potential in Eq.~(26) of Ref.~\cite{PengPLA2011}, it is found that the thermodynamic potential $\Omega$ is proportional to $(\hbar\omega)^{-3}$, where the trapping frequency is given by $\omega=(\omega_x\omega_y\omega_z)^{1/3}$. Therefore, for an extremely anisotropic trap with $\omega_z\rightarrow0$, the thermodynamic potential $\Omega\rightarrow\infty$, which is an unphysical result. Therefore, the method in Refs.~\cite{PengPLA2011,PengPRA2011} cannot be applied to the trap which is extremely anisotropic with $\omega_z\rightarrow0$.
However, the thermodynamic potential (Eq.~(\ref{omegah})) in our work can be used for the extremely anisotropic trap with $\omega_z\rightarrow0$.

\section{Summary}\label{4}

In this work, we have presented a theoretical study on the width of the $s$-wave CIR in a quasi-1D atomic gas under tunable transversely anisotropic confinement.
We find that the width of the CIR can be tuned by varying the transverse anisotropy.
Typically, we calculate the resonance position and width of the CIR for $^{40}$K and $^{133}$Cs atoms.
Furthermore, we investigate the normal-state thermodynamics of the quasi-1D atomic gases.
As two typical examples, we also calculate the interaction energy densities of $^{40}$K and $^{133}$Cs atomic gases across both the wide and narrow CIRs with different transverse anisotropy.
We find that the change in the width of the CIR can manifest itself in the position of the discontinuity in the interaction energy density, which can be probed experimentally~\cite{InteractionEnergy}.

\section*{Acknowledgements}

We thank Wei Yi, Wei Zhang, Xiaoling Cui, Ming Gong, Lijun Yang, Jingbo Wang, V. S. Melezhik, and P. Schmelcher for helpful discussions.
We also thank the referees for improving the quality of this manuscript.
This work is supported by the National Key R\&D Program (Grant No. 2016YFA0301700), the National Natural Science Foundation of China (Grant No. 11404106), and the ``Strategic Priority Research Program(B)'' of the Chinese Academy of Sciences (Grant No. XDB01030200). F.Q. acknowledges support from the Project funded by China Postdoctoral Science Foundation (Grant No. 2016M602011).
J.-S.P. acknowledges support from National Postdoctoral Program for Innovative Talents of China (Grant No. BX201700156).

\end{CJK*}
\end{document}